\documentclass[12pt,a4paper]{article}
\usepackage[margin=2.5cm]{geometry}
\usepackage{hyperref}
\hypersetup{colorlinks=true,
citecolor=ForestGreen, linkcolor=teal, urlcolor=teal}

\def\be{\begin{equation}}
\def\ee{\end{equation}}

\usepackage[usenames,dvipsnames]{xcolor}

\usepackage{tikz}
\usepackage{verbatim} 

\def\ba{\begin{eqnarray}}
\def\ea{\end{eqnarray}}
\def\nn{\nonumber}
\def\lb{\label}
\def\bb{\bibitem}
\def\dfrac{\displaystyle\frac}
\def\as{{\rm arcsinh}\,}
\def\ac{{\rm arccosh}\,}
\def\at{{\rm arctanh}\,}
\def\q{\bar{q}}
\begin{document}

\begin{titlepage}
\date{March 30, 2024}
\title{
\begin{flushright}\begin{small} LAPTH-015/24 \end{small} \end{flushright} \vspace{1cm}
Spherical symmetric solutions of\\ conformal Killing gravity: black holes, wormholes,\\ and sourceless cosmologies}



\author{G\'erard Cl\'ement$^a$\thanks{Email: gclement@lapth.cnrs.fr} and
Khireddine Nouicer$^{b}$\thanks{Email: khnouicer@univ-jijel.dz} \\ \\
$^a$ {\small LAPTh, Universit\'e Savoie Mont Blanc, CNRS, F-74940  Annecy, France} \\
$^b$ {\small LPTh, Department of Physics, University of Jijel,} \\
{\small BP 98, Ouled Aissa, Jijel 18000, Algeria}}

\maketitle

\begin{abstract}
The most general set of static and spherically symmetric solutions for conformal Killing gravity coupled to Maxwell fields is presented in closed form. These solutions, depending on six parameters, include non-asymptotically flat black holes or naked singularities, non-asymptotically flat traversable wormholes, and (possibly singularity-free) closed universes. We also consider the inverse problem, showing that the most general energy-momentum tensor generating a given static spherically symmetric metric depends on three parameters. Sourceless time-dependent isotropic solutions are also given. These solutions depending on the curvature, the cosmological constant, and a new integration constant $\alpha$, present a rich variety, including singularity-free eternal cosmologies and universes evolving symmetrically from a big bang to a big crunch within a finite lapse of time.
\end{abstract}

\end{titlepage}
\setcounter{page}{2}

\section{Introduction}\label{sec:intro}

In the pursuit of understanding the universe's accelerated expansion \cite{riess98,perlmutter99}, numerous alternatives have been proposed, both within the framework of general relativity and beyond. One such alternative involves introducing a new matter component termed dark energy, characterized by negative pressure. While the cosmological constant $\Lambda$, originally introduced by Einstein, remains the primary candidate for dark energy due to its alignment with observational data, challenges such as the fine-tuning  and coincidence problems persist \cite{weinberg89,zlatev99, Bull,Velten}. Recently, the DESI collaboration \cite{desicollaboration2024desi} has found evidence for dynamical dark energy in the form of quintessence. This finding implies that dark energy may not be a constant, necessitating a re-examination of the concordance model and offering new insights into the universe's expansion and ultimate fate. On the other hand, a different path follows the idea that general relativity fails to describe the universe at larger scales and hence it should be modified. These modifications often propose alterations to fundamental principles of physics, seeking to reconcile observational evidence with theoretical predictions \cite{skordis,saridakis}. Consequently, alternative models like scalar tensor theories \cite{bergmann,kase}, Einstein-\ae{}ther theory \cite{Jacobson1,Jacobson2}, and higher derivative theories of gravity like $f(R)$ theories \cite{Sotiriou,Nojiri} have been explored. Other kinds of modified theories are the modified teleparallel equivalent of general relativity, $f(T)$ gravity where the Ricci scalar is replaced by the torsion scalar, and  the theory grounded in Weyl geometry, termed $f(Q)$ gravity \cite{Aldrovandi}. The latter describes gravitational effects in terms of non-metricity, providing a broader framework than traditional Riemannian geometry. These theories offer different perspectives on gravity that may lead to new insights into cosmological and astrophysical phenomena.

Recently a new gravitational field theory was proposed by J. Harada \cite{harada2023a}, and subsequently termed ``conformal Killing gravity'' \cite{mantica2023}. By construction, all solutions of the Einstein equations with or without cosmological constant (which arises as an integration constant of Harada's field equations) are also solutions of this new theory. But, the field equations of conformal Killing gravity being differential equations of third order, it also yields solutions beyond those of cosmological general relativity. The application of conformal Killing gravity to cosmology has led to solutions that shed light on the transition from decelerating to accelerating expansion, offering an explanation for the current cosmic acceleration \cite{harada2023b}. Black hole solutions to conformal Killing gravity coupled to nonlinear electrodynamics have been explored in \cite{tarciso}.

Very recently, A. Barnes \cite{barnes2023a,barnes2023b} considered the field equations of conformal Killing gravity coupled to Maxwell fields and obtained static solutions in the form of infinite power series using the Frobenius method. In this work, we revisit this problem and derive the most general static and spherically symmetric solutions to conformal Killing gravity field equations sourced by Maxwell field in closed and compact form. The structure of this paper is outlined as follows. In the next section, we introduce the field equations of conformal Killing gravity, and derive the solutions. In Sec. \ref{sec:geom} we explore the geometric properties of the solutions, which contain black holes, wormholes, and singularity-free closed universes. In Sec. \ref{sec:Inv-pb}, we investigate the inverse problem, deriving the most general energy-momentum tensors which generate the metrics of Sec. \ref{sec:sol_sphero}. In Sec. \ref{sec:sourless-cosmo} we consider the class of time-dependent Friedmann-Lema\^{\i}tre-Robertson-Walker (FLRW) cosmologies solving the sourceless equations of conformal Killing gravity and discuss them  for flat, closed, and open space-times. Finally, we provide in  Sec. \ref{sec:discussion} a summary and discussion of our results.

\setcounter{equation}{0}
\section{Static spherically symmetric solutions with Maxwell source}\label{sec:sol_sphero}

The gravitational field equations proposed by Harada \cite{harada2023a} are expressed as follows:
\be\lb{ckg}
H_{\mu\nu\rho} = 8\pi G T_{\mu\nu\rho},
\ee
where the totally symmetric tensor $H_{\mu\nu\rho}$ is defined by
\begin{equation}
H_{\mu\nu\rho} \equiv \nabla_{\mu}R_{\nu\rho}+\nabla_{\nu}R_{\rho\mu}+\nabla_{\rho}R_{\mu\nu}
-\frac{1}{3}\left(g_{\nu\rho}\partial_{\mu}+g_{\rho\mu}\partial_{\nu}+g_{\mu\nu}\partial_{\rho}\right)R,
\end{equation}
with $R_{\mu\nu}$ the Ricci tensor, $R$ its trace, and $\nabla$ the covariant derivative. The components of this three-tensor are constrained by the Bianchi identity
\be\lb{bianchi}
g^{\nu\rho} H_{\mu\nu\rho} \equiv 0 .
\ee
The matter three-tensor $T_{\mu\nu\rho}$ is similarly defined as:
\begin{equation}\label{T3}
T_{\mu\nu\rho} \equiv \nabla_{\mu}T_{\nu\rho}+\nabla_{\nu}T_{\rho\mu}+\nabla_{\rho}T_{\mu\nu}
-\frac{1}{6}\left(g_{\nu\rho}\partial_{\mu}+g_{\rho\mu}\partial_{\nu}+g_{\mu\nu}\partial_{\rho}\right)T,
\end{equation}
with $T_{\mu\nu}$ the energy-momentum tensor, and $T$ its trace. The conservation of energy-momentum is embodied in the equation
\be\lb{conserv}
g^{\nu\rho} T_{\mu\nu\rho} = 0 .
\ee

It was shown in \cite{mantica2023} that the Harada field equations (\ref{ckg}) are equivalent to the Einstein equations
\be
R_{\mu\nu} = 8\pi G(T_{\mu\nu} + K_{\mu\nu})
\ee
modified by the addition to the matter energy-momentum tensor of a divergence- free ($\nabla_{\nu}K^{\mu\nu} = 0$) conformal Killing tensor $K_{\mu\nu}$ of trace $K$ satisfying the equation
\be
\nabla_{\mu}K_{\nu\rho}+\nabla_{\nu}K_{\rho\mu}+\nabla_{\rho}K_{\mu\nu} - \frac{1}{6}\left(g_{\nu\rho}\partial_{\mu}+g_{\rho\mu}\partial_{\nu}+g_{\mu\nu}\partial_{\rho}\right)K = 0.
\ee
This formulation will not be used in the following.

Let us now recall the derivation of the differential equations for static spherically symmetric metrics sourced by a Maxwell field \cite{barnes2023b}. The general static and spherically symmetric spacetime metric can be written as
\begin{equation}\lb{SSS}
ds^2 = -e^{2\nu(r)}dt^2 + e^{2\lambda(r)}dr^2 +r^2d\Omega^2
\end{equation}
($d\Omega^2=d\theta^2+\sin^2\theta\,d\varphi^2$). The monopole solution of the Maxwell equations in this geometry leads to the energy-momentum tensor components
\be\lb{monomax}
{T^t}_t = {T^r}_r = -{T^\theta}_\theta = -{T^\varphi}_\varphi = - \dfrac{q^2}{8\pi}\,r^{-4},
\ee
where $q^2$ is the sum of the squared monopole electric and magnetic charges. The resulting components of the three-tensor (\ref{T3})  are
\be
{T^t}_{tr} = - \dfrac13{T^r}_{rr} = \dfrac12{T^\theta}_{\theta r} =\dfrac12{T^\varphi}_{\varphi r} = - \dfrac{q^2}{2\pi}\,r^{-5}.
\ee

By virtue of the constraints (\ref{bianchi}) and (\ref{conserv}) the only independent equations (\ref{ckg}) are, for $G=1$,
\ba\lb{Etrr}
{H^t}_{tr} - 8\pi{T^t}_{tr} &\equiv& \dfrac13e^{-2\lambda}\left[\nu''' + (2\nu'-3\lambda')\nu'' - \nu'\lambda'' - 2(\nu'-\lambda')\nu'\lambda' \right.\nn\\
&& \left. + \dfrac2r(\nu''+2\lambda''-6\nu'^2-8\nu'\lambda'-4\lambda'^2) - \dfrac2{r^2}\nu' + \dfrac4{r^3}\right] \nn\\
&& - \dfrac4{3r^3} + \dfrac{4q^2}{r5} = 0,
\ea
where $'{} = d/dr$, and the combination
\be
{H^t}_{tr} - \dfrac13{H^r}_{rr} = g_{tt}\dfrac{d}{dr}\left[g_{tt}^{-1}({R^t}_t-{R^r}_r)\right] = -2e^{2\nu}\left[e^{-2(\nu+\lambda)} \dfrac{(\nu'+\lambda')}r\right]' = 0.
\ee
This last equation is solved by
\begin{equation}
e^{-2(\lambda(r) + \nu(r))} = c +d\, r^2,
\end{equation}
where $c$ and $d$ are integration constants. Using this equation to eliminate $\lambda(r)$ in terms of $\nu(r)$ from equation (\ref{Etrr}), and putting $e^{2\nu(r)} = y(r)$, we arrive at the master linear differential equation
\be\lb{master}
(c+d\,r^2)r^3y'''(r) + (-2c+d\,r^2)r^2y''(r) + (-2c-d\,r^2) ry'(r) + 8c y(r) - 8 + \frac{24 q^2}{r^2} = 0.
\ee

So the metric is
\be\lb{met}
ds^2 = -y(r)\,dt^2 + \dfrac{dr^2}{(c+d\,r^2)y(r)} + r^2\,d\Omega^2,
\ee
where $y(r)$ solves equation (\ref{master}). For this metric to have the Lorentzian signature, the product $g_{tt}g_{rr} = - (c+d\,r^2)$ must be negative definite. Three cases must be considered.

\underline{Case 1: $d=0$.}
The metric has the Lorentzian signature provided $c > 0$. By rescaling the time coordinate, one can fix $c=1$.
The solution of the master equation (\ref{master}) is \cite{harada2023a}
\begin{equation}\lb{harada}
y(r)= 1 - \frac{2m}{r} + \frac{q^2}{r^2} - \frac{\Lambda r^2}3 - \frac{\lambda r^4}5.
\end{equation}

\underline{Case 2: $d \neq0,\,c\neq0.$}
The metric has the Lorentzian signature in three subcases: 2a) $c > 0,\,d > 0$, with $r$ real; 2b) $c < 0,\,d > 0$, with $r^2 > -c/d$; 2c) $c > 0,\,d < 0$, with $r^2 < -c/d$. In these three cases, the modulus of $c$ can be set to $1$ by a rescaling of the time coordinate, so we can fix $c = \epsilon = \pm1$. Then, defining $\mu^2 = 1/|d|$, one may write $d = \eta/\mu^2$, where $\eta = \pm 1$, and $\mu$ is a constant with the dimension of a length. The generic solution for these three cases is
\ba\lb{sol2}
y(r) &=& \frac{q^2(2\eta r^2+\epsilon\mu^2)}{\mu^2r^2} + k_1\dfrac{\sqrt{\eta r^2 + \epsilon\mu^2}}r  + \epsilon \nn\\
&& + k_2\left[\dfrac{\sqrt{\eta r^2 + \epsilon\mu^2}}r\,\psi\left(\dfrac{r}\mu\right) - 1\right] + k_3 r^2,
\ea
where
\ba
a) \psi(x) =& \as{x}, \quad \epsilon = +1, \; \eta = +1, & \lb{psia}\\
b) \psi(x) =& \ac{x}, \quad \epsilon = -1, \; \eta = +1 & \quad (r^2 > \mu^2), \lb{psib}\\
c) \psi(x) =& \arcsin{x}, \quad \epsilon = +1, \; \eta = -1 & \quad (r^2 < \mu^2). \lb{psic}
\ea

Expanding the solutions 2a and 2c with $\epsilon = +1$ in Laurent series for the variable $r/\mu$, we recover the solutions given by Barnes (equation (20) of \cite{barnes2023b}):
\be
y(r) = 1 + (1+2d\,r^2)q^2/r^2 - 2mp_1(r) + \lambda p_2(r)r^4/5 - \Lambda r^2/3,
\ee
where $m = -k_1\mu/2$, $\lambda = -2k_2/3\mu^4$, $\Lambda = -\eta k_2/\mu^2 - 3k_3$, and the series $p_1(r)$ and $p_2(r)$ are summed up as
\ba
p_1(r) &=& \sqrt{1+ d\,r^2},\\
p_2(r) &=& -\dfrac{15}{2dr^4}\left[\dfrac{\mu\sqrt{1+ d\,r^2}}{r}\psi\left(\dfrac{r}\mu\right) - 1 - \dfrac{d\,r^2}3\right].
\ea
The solution 2b with $\epsilon = -1$, $\eta = +1$ was previously given by Barnes (equation (24) of \cite{barnes2023b}) for the special case $k_1=k_2=0$.

\underline{Case 3: $c=0$.}
In this case, the constant $d$ must be positive. Putting again $d = 1/\mu^2$, the solution of equation (\ref{master}) is \cite{barnes2023b}
\begin{equation}
y(r) = \frac{q^2\mu^2}{4r^4} - \frac{\mu^2}{2r^2} + k_1 + k_2\ln\left\vert\dfrac{r}\mu\right\vert - \frac{\Lambda r^2}3.
\end{equation}

\setcounter{equation}{0}
\section{Geometry}\label{sec:geom}

The coordinate singularities of the metric (\ref{met}) are at $r=0$ (possible curvature singularity), $y(r) = 0$ (horizons), $y(r) \to\infty$ (possible curvature singularity), and $c+d\,r^2=0$ (cases 2b and 2c). We will see in the following that similarly to the case of regular horizons, where maximal analytic extension may be carried out by transforming to Kruskal-like coordinates, the metric (\ref{met}) may be extended through the zeroes of $c+d\,r^2$.

\subsection{Case 1}

When $\lambda = 0$ the charged Harada solution (\ref{harada}) reduces to the well-known Reissner-Nordstr\"om-(anti-)de Sitter solution of general relativity. It was suggested by Harada \cite{harada2023a} that, when $\lambda \neq 0$, the new term in $r^4$ may become significant at large distances. However the Ricci scalar
\be
R = 4\Lambda + 6\lambda r^2
\ee
diverges on the sphere $r \to \infty$. Furthermore, the coordinate transformation $r = x^{-1}$ leads to the asymptotic form of the metric
\be
ds^2 \simeq \dfrac\lambda{5}\,x^{-4} dt^2 - \dfrac5\lambda\,dx^2 + x^{-2}\,d\Omega^2  \quad (x \to 0),
\ee
showing that this singular sphere is actually at finite geodesic distance for all geodesics if $\lambda > 0$, and for spacelike radial geodesics if $\lambda < 0$. If $m$, $\Lambda$ and $\lambda$ are negative, $y(r)$ is positive definite and the spacetime described by metric (\ref{met}) ends on the two naked singularities $r=0$ and $r\to\infty$. In the generic case there can be up to six horizons between these two singularities. If we assume for simplicity $q=\Lambda=0$ and $m>0$, then for $\lambda < 0$ the stationary domain is bounded on one side by a horizon shielding the Schwarzschild-like spacelike singularity $r=0$ and on the other side by the naked timelike singularity $r\to\infty$, while for $\lambda > 0$ two horizons $r = r_\pm$ hide two spacelike singularities.

\subsection{Case 2}

In this case the Ricci scalar is
\ba
\mu^2R &=& -6\eta\left[\dfrac{\epsilon q^2}{r^2} + \left(k_1+k_2\psi\left(\dfrac{r}\mu\right)\right)\dfrac{\sqrt{\eta r^2 + \epsilon\mu^2}}r + \epsilon + \dfrac{2\eta q^2}{\mu^2} + 3k_3 r^2\right] \nn\\
&&
+ 2\eta k_2 - 12\epsilon k_3\mu^2  .
\ea
The possible curvature singularities are at $r \to \infty$ and at the center $r = 0$. Similarly to case 1, the singularity $r \to \infty$ is at finite geodesic distance in subcases 2a or 2b unless $k_3 = 0$. If $k_3 = 0$ but $k_2 \neq 0$, the logarithmic divergence of the function $\psi(r/\mu)$ (an \as or \ac) at $r \to \infty$ in subcases 2a or 2b carries over to the curvature invariants. However, the coordinate transformation $r = \mu e^x$ leads for $d = 1/\mu^2$ to the asymptotic form of the metric
\be
ds^2 \simeq - k_2xdt^2 + \dfrac{dx^2}{k_2x} + \mu^2 e^{2x} d\Omega^2  \quad (x \to \infty),
\ee
showing that spacelike (for $k_2 > 0$) or timelike (for $k_2 < 0$) geodesics extend to $r \to \infty$. In other words, geodesics do not terminate at $r \to \infty$ if $k_3 = 0$ but $k_2 \neq 0$. The invariants also diverge for $r \to 0$, unless $q=0$ and $k_1=0$.

\subsubsection{Subcase 2a}

We choose $k_3 = 0$ in (\ref{sol2}) to avoid the curvature singularity at $r \to \infty$, and discuss first the simple case $k_2 = 0$. In the vacuum case $q=0$,
\be\lb{yav0}
y(r) = 1 + k_1\frac{\rho}r  \quad \left(\rho = \sqrt{r^2 + \mu^2}\right).
\ee
For $k_1 \le -1$, there is no stationary domain ($y(r)$ is negative definite). For $-1 < k_1 < 0$, the spacetime [\ref{met}) is a non-asymptotically flat black hole, with a single horizon shielding the singularity $r=0$. For $k_1 > 0$ the singularity is naked. For $k_1 = 0$, the singularity-free metric
\be
ds^2 = - dt^2 + \dfrac{dr^2}{1+r^2/{\mu^2}} + r^2\,d\Omega^2
\ee
is that of a negative density Einstein static universe \cite{barnes2023a}.

In the electromagnetic case, $q \neq0$, the lapse function is
\be\lb{yae0}
y(r) = \q^2\frac{\rho^2}{r^2} + k_1\frac{\rho}{r } + \q^2 + 1 \quad (\q = q/\mu).
\ee
For $k_1 \le -(1+ 2\q^2)$, $y(\infty)<0$ so that the central timelike singularity is not shielded by the cosmological horizon. For $-(1+2\q^2) < k_1 < -2|\q|\sqrt{1+\q^2}$  the spacetime is a black hole with two horizons and a central singularity. It becomes an extreme black hole for $ k_1 = -2|\q|\sqrt{1+\q^2}$, with a double horizon at $r = |q|$. For $k_1 > -2|\q|\sqrt{1+\q^2}$, the singularity is naked.

Consider now the general case $k_2 > 0$ (the sign being chosen so that $y(\infty) > 0$). The lapse function $y$ is now
\be
y(r) = y_0(r) + k_2z(r),
\ee
where $y_0(r)$ is the function (\ref{yav0}) or (\ref{yae0}), and $z(r) = (\rho/r)\as(r/\mu) - 1$ is a positive, steadily increasing function, such that $z(0) = z'(0) = 0$. It follows that the results obtained above for $k_2=0$ remain qualitatively valid (without a lower bound on $k_1$). For $q=0$ the spacetime is a black hole with a single horizon if $k_1 < 0$, it is geodesically complete if $- 1 < k_1 < 0$, and the singularity is naked if $k_1 > 0$. For $q\neq0$ the spacetime corresponds either to a black hole with a double horizon, to an extreme black hole, or to a naked singularity, depending on the parameter values.

\subsubsection{Subcase 2b}
\
We again choose $k_3 = 0$ to avoid the curvature singularity at  $r \to \infty$. In this case, because $r^2 > \mu^2$, the metric is best parameterized in terms of the radial coordinate $\rho = \sqrt{r^2 - \mu^2}$,
\be
ds^2 = - y(\rho)\,dt^2 + \dfrac{\mu^2 d\rho^2}{(\mu^2+\rho^2)y(\rho)} + (\mu^2+\rho^2)\,d\Omega^2,
\ee
with
\be
y(\rho) = \q^2 x^2 + k_1 x + \q^2 - 1 + k_2[x\,\at{x} - 1],\;\;  x = \dfrac\rho{\sqrt{\mu^2+\rho^2}}\;\; (-1 < x < 1).
\ee
While the radial coordinate $\rho$ has been defined to be positive, there is clearly no coordinate singularity for $\rho=0$, so that the spacetime should be analytically extended to $\rho<0$. The variable $\rho$ then takes its values in the whole real axis, so that there are two points at infinity $\rho \to \pm\infty$ ($x \to \pm1$), and a sphere of minimal area $4\pi\mu^2$ for $\rho=0$.

We first discuss the case $k_2 = 0$. The central singularity is absent, so that the spacetime will be geodesically complete after analytical extension. Because of the quadratic form of the metric function $y$, there are at most two horizons. The extremal values of $y(\rho)$ are
\be
y(\pm\infty) = 2\q^2 - 1 \pm k_1, \quad y_{\rm min} = \dfrac{4\q^2(\q^2-1) - k_1^2}{4\q^2}.
\ee
From these values we see that the range of the parameter $|k_1|$ can be divided in three segments.
\begin{enumerate}
\item $k_1^2 > (2\q^2 - 1)^2$. In this case, which includes the vacuum solution $q = 0$, the spacetime is a singularity-free black hole with one horizon.
\item $4\q^2(\q^2-1) < k_1^2 < (2\q^2 - 1)^2$ and $\q^2 > 1/2$ (for $\q^2 < 1/2$, $y(\rho)$ is negative definite so that there is no stationary domain). In this case, the spacetime is a singularity-free black hole with two horizons.
\item $k_1^2 < 4\q^2(\q^2-1)$. In this case, which occurs only if $\q^2 > 1$, $y(\rho)$ is positive definite so that the spacetime is a traversable wormhole, which is symmetrical if $k_1=0$.
    \end  {enumerate}

The existence of a traversable wormhole solution to conformal Killing gravity is a remarkable feature, since these are forbidden in general relativity by the null energy condition \cite{Morris:1988}. To show explicitly that this wormhole is traversable, let us write down the first integrated geodesic equation for timelike equatorial ($\theta=\pi/2$) geodesics, with $E$ and $L$ the energy and angular momentum of a unit mass test particle, as
\be
\dot{\rho}^2 + V(\rho) = 0,
\ee
where the effective potential is
\be
V(\rho) = \dfrac{\mu^2+\rho^2}{\mu^2}\left(y(\rho)-E^2\right) + \dfrac{L^2}{\mu^2}y(\rho).
\ee
The metric function $y(\rho)$ is bounded by $y(\rho)\le A$, where $A=2\q^2-1+|k_1|$ (the largest of the two values of $y(\rho)$ at infinity). It then follows that the first term of $V(\rho)$ is negative with the upper bound $A-E^2$ provided $E^2 > A$, while the positive second term has the upper bound $L^2A/\mu^2$. So the effective potential $V(\rho)$ is negative definite in the whole real range of the radial coordinate $\rho$ if the energy is large enough, $E^2 > A(1+L^2/\mu^2)$.

If $k_2 >  0$, the metric function $y$ diverges for $\rho \to \pm\infty$ as $y \simeq k_2 \ln|2\rho/\mu|$. Accordingly, there is necessarily an even number of horizons. It follows from $y(0) = \q^2 - 1 - k_2$ that, if
\be
\q^2 < 1 + k_2
\ee
(including the vacuum solution $q = 0$), the spacetime is a singularity-free black hole with two (or possibly more) horizons. It can be a traversable wormhole if $\q^2 > 1 + k_2$. In the case $k_1 = 0$, $y(\rho)$ is even in $\rho$ with a positive minimum at $\rho=0$ if
\be
\q^2 > 1 + k_2,
\ee
leading to a symmetrical traversable wormhole.

\subsubsection{Subcase 2c}

This case is complementary to the preceding, with $r^2 < \mu^2$. The metric is, in terms of the radial coordinate $\psi=\arcsin(r/\mu)$ ($0 < \psi < \pi/2$),
\be
ds^2 = - y(\psi)\,dt^2 + \mu^2\left[\dfrac{d\psi^2}{y(\psi)} + \sin^2\psi\,d\Omega^2\right],
\ee
with
\be
y(\psi) = \q^2\cot^2\psi + k_1\cot\psi + 1 - \q^2 + k_2(\psi\cot\psi - 1) + \mu^2k_3\sin^2\psi.
\ee
If $q$ and/or $k_1$ are different from $0$, the metric has a curvature singularity at the center $\psi = 0$. On the other hand, there is no curvature singularity at $\psi=\pi/2$, so that the range of $\psi$ may be extended (possibly with Kruskal-like analytical continuation through the zero(es) of $y(\psi)$) to $0 < \psi < \pi$, $\psi=\pi$ being again a curvature singularity.

If $q=k_1=0$, the metric becomes regular at $\psi = 0$, where $y(0) =  1$. It follows that the range of $\psi$ may again be extended to $-\pi < \psi < \pi$, with again curvature singularities at the two ends $\psi = \pm\pi$ of this interval.

If furthermore $k_2=0$, the metric function reduces to
\be
y(\psi) = 1 - \dfrac{\Lambda\mu^2}3\,\sin^2\psi.
\ee
The singularity-free spacetime is then geodesically complete (possibly after Kruskal continuation). It is stationary everywhere if $\Lambda < 3/\mu^2$. This case corresponds to a regular, closed static universe, with finite volume $V = 2\pi^2\mu^3$. In the special subcase $k_3 = 0$
($\Lambda=0$), this is the Einstein static universe.

\subsection{Case 3}

The Ricci scalar for this case is
\be
\mu^2R = - \dfrac{3q^2\mu^2}{2r^4} + \dfrac{3\mu^2}{r^2} - 4k_2 - 6\left(k_1+k_2\ln\left\vert\dfrac{r}\mu\right\vert\right) + 6\Lambda r^2.
\ee

As in subcases 2a and 2b, we choose $\Lambda = 0$ to avoid a curvature singularity for $r \to \infty$. The metric function $y$ is
\be
y(r) = \frac{\q^2\mu^4}{4r^4} - \frac{\mu^2}{2r^2} + k_1 + k_2\ln\left\vert\dfrac{r}\mu\right\vert,
\ee
with $k_2>0$ or ($k_2=0$ and $k_1>0$) for the solution to be stationary at $r\to\infty$. The central singularity $r=0$ is always present. In the vacuum case $q=0$, the spacetime is a black hole with a single horizon. In the electromagnetic case $q \neq0$, it can correspond either to a black hole with two horizons, an extreme black hole, or a naked singularity, depending on the parameter values.

\setcounter{equation}{0}
\section{The inverse problem}\label{sec:Inv-pb}

Let us recall that any solution of Einstein's field equations is, by construction, a solution of the equations of conformal Killing gravity. The static Einstein universe is a solution of general relativity sourced by a perfect fluid with constant density $\rho$ and opposite pressure $p$. We have seen that the static Einstein universe is also a vacuum solution ($T_{\mu\nu}=0$) of conformal Killing gravity, which implies that its perfect fluid source must satisfy the condition $T_{\mu\nu\rho}=0$, and also generate the other vacuum solutions of Section \ref{sec:sol_sphero}. This prompts us to examine the inverse problem: What is the most general energy-momentum tensor generating a given metric (for instance the Schwarzschild solution)?

We will not attempt here to answer this question in all its generality, but restrict our investigation to the case of a static spherically symmetric matter source \footnote{The spherical symmetry of $T_{\mu\nu\rho}$ does not necessarily imply the spherical symmetry of $T_{\mu\nu}$. }. This is characterized by a diagonal energy-momentum tensor $T_{\mu\nu}$, which may be thought of as a static anisotropic fluid, with components
\be\lb{TS}
T_\mu^\nu = {\rm diag}\left(-\rho, p-2\Pi, p+\Pi, p+\Pi\right).
\ee

From (\ref{TS}) we obtain the components of the tensor $T_{\mu\nu\rho}$
\ba\lb{TTS}
T_{100} &\equiv& e^{2\nu}\left[\frac{5\rho'}{6} + \frac{p'}{2} - 2(\rho+p-2\Pi)\nu'\right], \nn\\
T_{111} &\equiv& 3e^{2\lambda}\left[\frac{\rho'}{6} + \frac{p'}{2} - 2\Pi'\right], \nn\\
T_{122} &\equiv& r^{2}\left[\frac{\rho'}{6} + \frac{p'}{2} + \Pi' - \frac{6\Pi}{r}\right], \;\; T_{133} = \sin^2\theta T_{122},
\ea
satisfying the conservation law (\ref{conserv}), which reads here
\be\lb{COS}
p' - 2\Pi' + (\rho+p-2\Pi)\nu' - \dfrac{6\Pi}r = 0.
\ee

The most general energy-momentum tensor generating the vacuum ($Q=0$) spacetime metrics of section 3 is such that the components (\ref{TTS}) all vanish. We first consider the combination
\be
- e^{-2\lambda}T_{111} + 3r^{-2}T_{122} \equiv 9\left(\Pi' - 2\dfrac{\Pi}r\right) = 0,
\ee
which is solved by $\Pi = (\alpha/3)r^2$, with $\alpha$ an integration constant. Inserting this in the equation $T_{111} = 0$, and integrating, we obtain
\be
p = \dfrac13(-\rho + 2\beta + 4\alpha r^2),
\ee
with $\beta$ a new integration constant. Finally, solving the equation $T_{100} = 0$ leads to the characteristic functions of the most general energy-momentum tensor (\ref{TS}) generating a given $Q=0$ spacetime metric (\ref{met}):
\be\lb{inverse}
 \rho(r) = \gamma y(r) - \alpha r^2 - \beta, \; p(r) = \dfrac13[-\gamma y(r) + 5\alpha r^2 + 3\beta], \; \Pi(r) = \dfrac{\alpha}3 r^2,
\ee
where $\gamma$ is a third integration constant, and $y(r)$ is the metric function for the spacetime under consideration. For instance, the most general energy-momentum tensor generating the Schwarzschild metric of mass $M$ in conformal Killing gravity corresponds to
\be
\rho_S(r) = \gamma(1-2M/r) - \alpha r^2 - \beta, \; p_S(r) = \dfrac13[-\gamma(1-2M/r) + 5\alpha r^2 + 3\beta], \; \Pi_S(r) = \dfrac{\alpha}3 r^2.
\ee
For $\alpha=\gamma=0$, this reduces to the energy-momentum generating the Einstein static universe in general relativity.

\setcounter{equation}{0}
\section{Time-dependent vacuum solutions}\label{sec:sourless-cosmo}

The occurrence of the Einstein static universe as a sourceless ($T_{\mu\nu}=0$) solution of conformal Killing gravity suggests that this theory might admit other sourceless cosmological solutions. Let us first recall the derivation of the Friedmann-like equation given in \cite{harada2023a}. Making the FLRW metric ansatz
\be\lb{FLRW}
ds^2 = - dt^2 + a^2(t)\left(\dfrac{dr^2}{1-kr^2} + r^2 d\Omega^2\right)
\ee
for an isotropic and spatially homogeneous universe, and assuming that the matter source is a perfect fluid with energy-momentum tensor
\be
T_\mu^\nu = {\rm diag}\left(-\rho, p, p, p\right)
\ee
satisfying the conservation law
\be\lb{conscosm}
\dot\rho + 3(\rho+p)\dfrac{\dot a}a = 0
\ee
(where $\dot{} = d/dt$), Harada \cite{harada2023a} showed that the equations of conformal Killing gravity may be first integrated to the second order equation
\be\lb{friedhara}
-\dfrac{\ddot{a}}{a} + \dfrac{2\dot{a}^2}{a^2} + \dfrac{2k}{a^2} - \dfrac\Lambda3 = \dfrac{4\pi G}3(5\rho+3p),
\ee
where the integration constant $\Lambda$ may be identified with the cosmological constant.

A first solution of this equation is the static solution $\dot{a}=0$, provided the density $\rho$ and pressure $p$ are time-independent. Then, rescaling the radial coordinate $r$ so that $a=1$, the FLRW metric (\ref{FLRW}) is of the form (\ref{met}) with $y(r) = 1$, $c=1$, $d = -k$, corresponding to the Einstein static universe. In the sourceless case ($\rho=p=0$), equation (\ref{friedhara}) gives $d = - \Lambda/6$. Note that the identification of the cosmological constant with an integration constant depends on which differential equations are integrated, the space-dependent equations for the functions $\nu(r)$ and $\lambda(r)$ in (\ref{SSS}), or the time-dependent equation (31) of \cite{harada2023a}.

Assuming now $\dot{a}\neq0$, let us now integrate again this equation as follows. Putting $a = b^{-1}$ and using the conservation law (\ref{conscosm}),
equation (\ref{friedhara}) may be rewritten as
\be
\dfrac{\ddot{b}}{b^3} + 2k  - \dfrac\Lambda{3b^2} - \dfrac{4\pi G\dot\rho}{3b\dot{b}} - \dfrac{8\pi G\rho}{3b^2} = 0,
\ee
which is first integrated by
\be
\dot{b}^2 + kb^4  - \dfrac{\Lambda + 8\pi G\rho}3b^2 = \alpha ,
\ee
where $\alpha$ is an integration constant, or
\be\lb{friedgen}
\dfrac{\dot{a}^2}{a^2} = \dfrac{8\pi G\rho}{3} + \dfrac{\Lambda}3 - \dfrac{k}{a^2} +  \alpha a^2.
\ee
This equation was previously obtained through an alternative method in \cite{mantica2023} (our constant $\alpha$ is equivalent to their constant $-C/6$). If $\alpha=0$ we recover the usual Friedmann equation. Thus, the integration constant $\alpha$ measures the deviation of conformal Killing gravity from cosmological general relativity \cite{mantica2023}.

A more direct interpretation of this constant has been proposed in \cite{harada2023b}. Consider the cosmological Einstein equations
\be
G_{\mu\nu} + \Lambda g_{\mu\nu} = 8\pi G(T_{\mu\nu} + T_{{\rm eff}\mu\nu}),
\ee
where $T_{\mu\nu}$ is the sum of the energy-momentum tensors for matter and radiation, and $T_{{\rm eff}\mu\nu}$ is the energy-momentum tensor for a hypothetical matter component, which Harada identified with effective dark energy, with density $\rho_{\rm eff}$ and pressure $p_{\rm eff}$ satisfying the equation of state
\be\lb{statealpha}
5\rho_{\rm eff} + 3p_{\rm eff}  = 0.
\ee
For this matter component, the integration of the conservation law (\ref{conscosm}) leads to the solution $\rho_{\rm eff}(t) = Ca^2(t)$ (with $C$ an integration constant). Thus, the contribution of this effective dark energy to the general-relativistic Friedmann equation
\be
\dfrac{\dot{a}^2}{a^2} = \dfrac{8\pi G\rho}{3} + \dfrac{8\pi G\rho_{\rm eff}}{3} + \dfrac{\Lambda}3 - \dfrac{k}{a^2}
\ee
can be identified with the term $\alpha a^2$ in (\ref{friedgen}),  if $C=\alpha/8\pi G$. So, according to Harada, this term can account for an effective dark energy which appears in the theory without needing to be introduced in the physical energy-momentum tensor $T_{\mu\nu}$.

In vacuum ($\rho = p = 0$), equation (\ref{friedgen}) reduces to
\be\lb{friedfirst}
\dot{a}^2 + V(a) = 0, \quad V(a) =  - \alpha a^4 - \dfrac{\Lambda}3a^2 +k.
\ee
The general solution to this equation can be written in terms of Jacobi functions, but it is enough to discuss the behavior of the solutions depending on the relative values of the integration constants $k$, $\Lambda$, $\alpha$, and on the sign of the discriminant of $V(a)$, which is
\be
\Delta = \dfrac{\Lambda^2}9 + 4k\alpha.
\ee
Let us discuss the two cases $\alpha>0$, and $\alpha<0$.

\subsection{\underline{$\alpha>0$}}
For $\Delta > 0$, the effective potential $V(a)$ has two real roots $a_\pm^2 = (-\Lambda/3 \pm \sqrt{\Delta})/2\alpha$, with $a_+^2 > 0$ if $\Lambda<0$ or $k>0$, and $a_-^2 > 0$ if $\Lambda<0$ and $k<0$. The scale factor $a(t)$ can vary from $a_+$ to infinity, or from $0$ to $a_-$, if $a_+$ or $a_-$ are real. If not, $a(t)$ varies from zero to infinity. Near infinity, $a(t)$ behaves as
\be
a \simeq \pm\dfrac1{\sqrt\alpha}\,\dfrac1{t-t_0} \quad (t \to t_0),
\ee
for some finite $t_0$.

For \underline{$k>0$} and all values of $\Lambda$, the evolution is time-symmetrical, with a minimum scale function $a(t)$ at $t=0$ and two asymptotes at $t = \pm t_0$.

For \underline{$k=0$}, equation (\ref{friedfirst}) can be transformed by putting $a(t)= 1/b(t)$ to $\dot{b}^2  - (\Lambda/3)b^2 = \alpha$, which can be solved in terms of elementary functions:
\ba
a = & \pm\dfrac{\sqrt{\Lambda/3\alpha}}{\sinh\left(\sqrt{\Lambda/3} t\right)} \quad & (\Lambda > 0), \nn\\
a = & \pm\dfrac1{\sqrt{\alpha}t} \quad & (\Lambda = 0), \\
a = & \pm\dfrac{\sqrt{-\Lambda/3\alpha}}{\sin\left(\sqrt{-\Lambda/3} t\right)} \quad & (\Lambda < 0) \nn
\ea
(up to a time translation). If $\Lambda\ge 0$ ($a_+=0$) the scale factor asymptotes to zero for $t = \pm\infty$ and to infinity for $t=0$. If $\Lambda < 0$ ($a_+=\sqrt{-\Lambda/3\alpha}$), the evolution is the same as for $k>0$.

For \underline{$k< 0$}, $\Lambda > -6\sqrt{-k\alpha}$, $V(a)$ is negative definite, so $a(t)$ varies from zero to infinity, with the exact solution if $\Lambda = 6\sqrt{-k\alpha}$,
\be
a = \sqrt{\Lambda/6\alpha}\vert\tan(\sqrt{\Lambda/6}t)\vert.
\ee
If $\Lambda = - 6\sqrt{-k\alpha}$, we have again the two exact solutions
\be
a = \sqrt{-\Lambda/6\alpha}\vert\tanh(\sqrt{\Lambda/6}t)\vert, \;\; {\rm or} \;\; a = \sqrt{-\Lambda/6\alpha}\vert\coth(\sqrt{\Lambda/6}t)\vert.
\ee
If $\Lambda < - 6\sqrt{-k\alpha}$, the two roots are real, so that the evolution is time symmetrical, with either the universe contracting to a minimum radius at $a = a_+$ and then expanding again to infinity in a finite time interval $2t_0$; or blowing up from a big bang at $t=-t_0$ to a maximum radius at $a = a_-$ and then recontracting to a big crunch at $t=t_0$.

\subsection{\underline{$\alpha<0$}}
This is possible only if $\Delta > 0$. The squared scale factor $a^2(t)$ must remain between the two roots $a_\pm^2 = (\Lambda/3 \pm \sqrt{\Delta})/(-2\alpha)$, implying $a_+^2 > 0$.

For \underline{$k>0$}, $a_+^2$ is positive only if $\Lambda > 6\sqrt{-k\alpha}$. In this case $a_-^2$ is also positive, leading to an oscillating cosmology, with the scale factor varying periodically between $a_-$ and $a_+$.

For \underline{$k=0$}, $\Lambda>0$, we have the exact solution
\be
a = \dfrac{\sqrt{\Lambda/(-3\alpha)}}{\cosh\left(\sqrt{\Lambda/3} t\right)}.
\ee

For \underline{$k< 0$}, $a_+^2$ is positive and $a_-^2$ is negative for all values of $\Lambda$. The universe evolves symmetrically from a  big bang to a big crunch in a finite lapse of time.

\setcounter{equation}{0}
\section{Discussion}\label{sec:discussion}

We have obtained all the static spherically symmetric solutions, depending on six integration constants, of conformal Killing gravity with Maxwell source. These solutions fall in three classes, according to the values of the integration constants $c$ and $d$. The Schwarzschild-like class $d=0$ was previously given in \cite{harada2023a}. The solutions of this class depend on the mass $m$, electric or magnetic charge $q$, cosmological constant $\Lambda$, and Harada's integration constant $\lambda$. However, when $\lambda\neq0$, the metric is singular on the spheres $r\to\infty$, which are actually at finite geodesic distance. When $\lambda = 0$ the solution reduces to the well-known Reissner-Nordstr\"om-(anti-)de Sitter solution. The solutions of the class $c=0$, previously given in \cite{barnes2023b}, also present for $r\to\infty$ a naked singularity at finite geodesic distance, unless the cosmological constant $\Lambda$ (an integration constant in this theory) vanishes. The spacetimes with $c=0,\Lambda=0$ are not asymptotically flat, and can correspond to black holes or naked singularities, depending on the parameter values.

The class of solutions with non-zero integration constants $c$ and $d=\pm1/\mu^2$  is the richest. These solutions are given here for the first time in closed form, enabling an analysis of the geometry of these spacetimes. Depending on the signs of $c$ and $d$, these solutions can correspond to non-asymptotically flat black holes or naked singularities, to non-asymptotically flat traversable wormholes, or to (possibly singularity-free) closed universes. The traversable wormhole solutions occur only for overcharged solutions, $q^2 > \mu^2$. We can recall here that overcharged Reissner-Nordstr\"om-NUT solutions to the Einstein-Maxwell equations have been shown to correspond to traversable wormholes \cite{clement2016}. However these where actually generated by Misner-string line sources, violating the null energy condition \cite{clement2023}. Remarkably, the NUT-less wormhole solutions of conformal Killing gravity found here do not require the presence of exotic matter sources, the null energy condition being clearly satisfied by the monopole Maxwell configurations (\ref{monomax}).

We have also examined the inverse problem. Because the field equations of conformal Killing gravity involve only the covariant derivatives of the matter energy-momentum tensor, this cannot be uniquely determined from the knowledge of the spacetime metric. We have found that the most general energy-momentum tensor generating a given static spherically symmetric metric, for instance the Schwarzschild metric, depends on three parameters $\alpha$, $\beta$ and $\gamma$, see equation (\ref{inverse}). Moreover, all the energy-momentum tensors of (\ref{inverse}) with $\gamma=0$ are physically equivalent, in the sense that they generate all the vacuum ($\alpha=\beta=\gamma=0$) solutions of section \ref{sec:sol_sphero}.

The existence of singularity-free closed universe vacuum solutions such as the Einstein static universe prompted us to examine the possibility of other sourceless cosmological solutions to this theory. We found again a rich variety of time-dependent FLRW solutions depending on the curvature $k$, the cosmological constant $\Lambda$, and a new integration constant $\alpha$, the value $\alpha=0$ corresponding to cosmological general relativity. These solutions range from universes with a finite timeline for $\alpha>0$ to eternal universes for $\alpha<0$, and from singularity-free cosmologies for $k\ge0$ (including oscillating cosmologies for $\alpha<0,k\ge0$) to universes with a big bang and/or a big crunch for $k<0$.

The FLRW solutions we have found presumably do not exhaust the class of all time-dependent spherically symmetric sourceless solutions to conformal Killing gravity, which could arise from the metric ansatz
\begin{equation}\lb{birk}
ds^2 = -e^{2\nu(r,t)}dt^2 + e^{2\lambda(r,t)}dr^2 +r^2d\Omega^2
\end{equation}
more general than (\ref{FLRW}). At any rate, it is clear that the Birkhoff theorem is not valid for conformal Killing gravity. there being at least a three-parameter manifold of time-dependent spherically symmetric solutions.

In conclusion, our analysis has shown that the third-order differential field equations of conformal Killing gravity admit a rich variety of solutions. Further investigations into the potentialities of this new theory are necessary in order to ascertain its power as a substitute for general relativity.

\section*{Acknowledgments}

One of us (K.N.) thanks the LAPTh Annecy, France, where this work was initiated, for warm hospitality and the University of Jijel, Algeria, for financial support. We are grateful to J. Harada and A. Barnes for valuable comments.



\end{document}